\documentclass[sigconf, screen, nonacm]{acmart}
\setcopyright{none}
\keywords{Audio Classification, Transferability, Machine Learning, Causality}

\ccsdesc[500]{Computing methodologies~Causal reasoning and diagnostics}

\usepackage{comment}

\usepackage{multirow}
\usepackage{tikz}
\usetikzlibrary{calc, positioning, fit, shadows, backgrounds, arrows.meta}
\usepackage{dsfont}
\usepackage{pgf}
\usepackage{pgfplots}
\pgfplotsset{width=5cm,compat=1.9}
\pgfplotsset{every tick label/.append style={font=\tiny}}
\usepackage{amsmath}
\usepackage{amsthm}
\usepackage{cleveref}

\usepackage{pifont}
\usepackage{xspace}

\newcommand{\xai}{XAI\xspace}

\newcommand{\commentout}[1]{}

\newcommand{\ie}{i.e.\xspace}

\theoremstyle{plain}
\newtheorem{theorem}{Theorem}[section]
\newtheorem{proposition}[theorem]{Proposition}
\newtheorem{lemma}[theorem]{Lemma}

\theoremstyle{definition}
\newtheorem{definition}{Definition}

\theoremstyle{remark}

\theoremstyle{plain}

\newcommand{\asvspoof}{ASVSpoof2019\xspace}
\newcommand{\ritw}{ITW\xspace}
\newcommand{\freqrex}{\textsc{f}\textnormal{req}\textsc{r}\textnormal{e}\textsc{x}\xspace}
\newcommand{\fft}{$\mathcal{F}$\xspace}

\newcommand{\gtzan}{GTZAN\xspace}

\newcommand{\ravdess}{RAVDESS\xspace}

\newcommand{\sanchit}{MC$_1$\xspace}
\newcommand{\atuaans}{MC$_2$\xspace}
\newcommand{\pedro}{MC$_3$\xspace}
\newcommand{\whisper}{MC$_4$\xspace}
\newcommand{\AST}{MC$_5$\xspace}

\newcommand{\wiam}{VC$_1$\xspace}
\newcommand{\hondage}{VC$_2$\xspace}
\newcommand{\hubert}{VC$_3$\xspace}
\newcommand{\pollner}{VC$_4$\xspace}
\newcommand{\firdhokk}{VC$_5$\xspace}

\newcommand{\gust}{SP$_1$\xspace}
\newcommand{\mo}{SP$_2$\xspace}
\newcommand{\melody}{SP$_3$\xspace}

\newcommand{\U}{\mathcal{U}}

\newcommand{\cF}{\mathcal{F}}
\newcommand{\V}{\mathcal{V}}

\newcommand{\sat}{\models}

\newcommand{\lem}{\begin{lemma}}
\newcommand{\elem}{\end{lemma}}
\newcommand{\pro}{\begin{proposition}}
\newcommand{\epro}{\end{proposition}}

\newcommand{\dfn}{\begin{definition}}
\newcommand{\edfn}{\end{definition}}

\newcommand{\xam}{\begin{example}}
\newcommand{\exam}{\end{example}}

\usepackage{algorithm}

\usepackage{booktabs}
\usepackage{xcolor,colortbl}
\usepackage{longtable}
\usepackage{subcaption}
\usepackage{pgfplots}
\usepackage{pgfplotstable}
\usepackage{paralist}
\usepgfplotslibrary{fillbetween}
\pgfplotsset{compat=1.16}

\newcommand{\?}{\stackrel{?}{=}}

\begin{document}

\title{If It's Good Enough for You, It's Good Enough for Me: Transferability of Audio Sufficiencies across Models}

\author{David A. Kelly}
\email{david.a.kelly@kcl.ac.uk}
\orcid{0000-0002-5368-6769}

\author{Hana Chockler}
\orcid{0000-0003-1219-0713}
\email{hana.chockler@kcl.ac.uk}
\affiliation{%
  \institution{King's College London}
  \country{UK}
}

\begin{abstract}

In order to gain fresh insights about the information processing characteristics of different audio classification
models, we propose \emph{transferability analysis}. Given a minimal, sufficient signal for a classification on a model
$f$, transferability analysis asks whether other models accept this minimal signal as having the same classification  
as it did on $f$. 

We define what it means for a sufficient signal to be transferable and perform a large study over $3$ different
classification tasks: music genre, emotion recognition and deepfake detection. We find that transferability rates vary
depending on the task, with sufficient signals for music genre being transferable $\approx26\%$ of the time. The other
tasks reveal much higher variance in transferability and reveal that some models, in particular on deepfake detection, 
have different transferability behavior. We call these models `flat-earther' models.

We investigate deepfake audio in more depth, and show that transferability analysis also allows to us to discover
information theoretic differences between the models which are not captured by the more familiar metrics of accuracy and
precision. 



\end{abstract}

\maketitle

\section{Introduction}

Audio content analysis has the goal of building systems which successfully process acoustic
environments~\cite{lerch2022introduction}. Among the many common use cases are emotion and music genre classification
and spoof voice recognition. With the rise of generative AI, the problem of audio deepfakes, in particular, is also
becoming ever more important. These tasks are increasingly tackled using deep learning models. The problem with such
large, complex models is that their decision process is frequently opaque. Sturm, in~\cite{sturm2014simple}, likens
these models to the $19^{th}$ century horse, Clever Hans, which appeared to solve complex mathematical problems but was
in fact just reading unconscious signals from the questioner. Clever Hans was given lots of information---the problem and
visual cues combined---but only used the visual cues. For Hans, the visual cues were sufficient for solving the problem.

Clever Hans was unique: no other horses at the time exhibited similar abilities. Deep learning models, however, are not
unique. This raises the question: can the same signal be used for multiple different `horses', or is each horse using a
different part of the signal? Indeed, if each horse is correctly solving the problem, we would expect close to $100\%$
agreement from each horse. If, however, each horse is only reading some particular visual cue, then there is no reason
to expect that the horses all do equally well: there is no reason to assume they are all using the same visual cues. We
call this problem \emph{transferability analysis}.

\begin{figure*}[t]
  \includegraphics{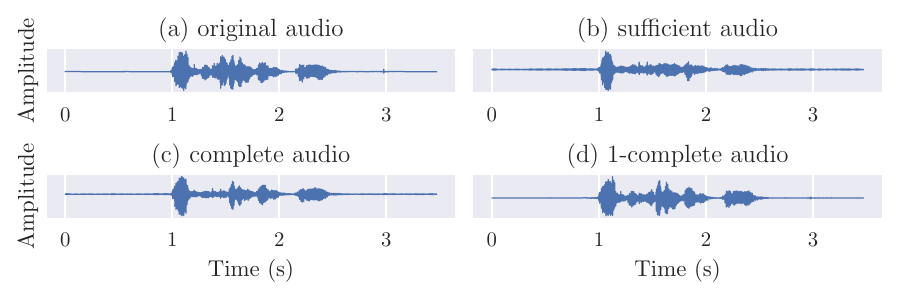}
    \caption{The partitioning of `disgust' (a) into sufficient signal (b), sufficient and necessary (complete) signal
    (c) and $1$-complete signal (d). A $1$-complete signal is sufficient and necessary for `disgust' and also has the same
    confidence as the original signal (a). We investigate how transferable these signals and classifications are
    across different models.}%
    \label{fig:disgust}
\end{figure*}

In this paper, we present the first study of transferability analysis. We find the cues that each model uses and test
other models to see whether they interpret the same cues in the same way. We use
\freqrex~\cite{kelly2026iguessthatsblues} to discover \emph{sufficient} and \emph{complete} signals for our
investigation. A sufficient signal is a minimal signal required for a classification, and a complete signal is a minimal
signal that is both sufficient and necessary for a classification~\cite{kelly2025causal}. \freqrex uses \emph{actual
causality} to perform a frequency domain analysis for a given signal $s$ and model $f$ (see~\Cref{fig:disgust}). The
relevant background into actual causality and \freqrex is presented in~\Cref{sec:background}.

In tabular data, one might reasonably expect that models all learn that the same features are important for any
particular class, though they may disagree in how much attention to pay to each feature. On higher dimensional data,
such as images, there is no reason to assume models all learn the same set of features to classify an image. Indeed, the size
and placement of minimal, sufficient pixel sets (MSPSs) used by different models suggests that pixel sets do not
transfer particularly well~\cite{kelly2025big}. These are suppositions: we are not aware of any studies into this issue.
In particular, it is completely unknown whether minimal audio sufficiencies are transferable across models.

We investigate transferability over $3$ different audio tasks: i) voice emotion classification; ii) music genre
classification; and iii) spoof audio classification. We identify and discuss `flat-earther' models: models whose
sufficiencies are not accepted by other models and which, in turn, do not accept the sufficiencies of others. In effect,
they do not listen to their peers, nor are they listened to. These models do not otherwise appear different,
having similar size, architectures and broadly similar performance. 

Deepfake spoof data receives special attention (\Cref{sec:spoof}). We show that is it possible to create composite `real' and
`fake' signals (\Cref{subsec:spoof_comp}) and that these have different spectral entropy and power spectral density
characteristics (\Cref{subsec:entropy}). Further, we perform an investigation with audio speech recognition
(ASR) techniques to assess the semantic fidelity of sufficient and complete signals (\Cref{subsec:asr}).

All of our code, data and results are available in the supplementary material.

\section{Background}\label{sec:background}

In what follows, we briefly introduce the relevant concepts from digital signal processing (DSP) and the theory of
actual causality. The reader is referred to~\cite{Hal19} for a more in-depth overview. Roughly speaking, actual causality
is suited to showing the causes of a particular event that occurred, rather than as a tool for predicting events, as is more common in
Pearl-style Bayesian causality. We also present relevant definitions for sufficiency and completeness as defined for
audio in~\cite{kelly2026guess}. We note that the prevalent framework for causality, which the reader is more likely to
be familiar with, is that of \emph{type causality}~\cite{pearl}. This framework is concerned with deriving
predictive---often probabilistic---causal models, based on statistical analysis of priors. This framework is not
suitable for our objective of analyzing specific instances. In contrast, the study of \emph{actual
causality}~\cite{Hal19} is backward-looking, and can help us derive \emph{causal explanations} underlying specific
events in the past.

\textbf{``Explanation'' and User Studies}: It has become the norm in \xai literature to assume that explanations ought to be aligned with human expectations. ~\cite{Hal19}, however, uses the word ``explanation'' to define a certain relationship between actual causes and causal settings (\Cref{defn:simple-exp}). It does not imply human \emph{interpretability}. This is how we use the word explanation in this paper.
Causal explanations have much more in common with work on logic-based explanations than they do with the
mainstream of \xai literature~\cite{INMS19,kelly2025causal}. In actual causality, a set of variables is an explanation if it satisfies certain requirements, none of which is human intuition. We are interested in signals minimal for a model; whether they align with human intuition is not our topic of concern. For this reason, we also do not pursue user study.

Even a sufficient signal---or set of pixels---may not be explanatory of a model's process. For example, in computer vision, it is not
usually possible to link a set of pixels, sufficient for a class, to a prediction instance $f(x)$ and say,
categorically, it was \emph{these} pixels which generated $f(x)$ due to the multiplicity of MSPSs~\cite{CKK25}, though
some approaches may be less likely to be out-of-distribution (OOD)~\cite{chanchal2025activation}.

\freqrex uses a frequency-based analysis to discover different subsets of the signal. For this, we require the Fourier
Transform to be well-defined over the signal. Most importantly, we assume the signal to be periodic.

\paragraph{Fourier Transform \fft.} For a given signal $f(x)$, 
\[ \mathcal{F}(\xi) = \int_{-\infty}^{\infty} f(x) e^{- 2 \pi \xi x} dx \, \forall \xi \in \mathbb{R} \] 
Multiplication (denoted $\cdot$) and convolution (denoted $*$) transform into each other, i.e. $\mathcal{F}[ s * r] =
\mathcal{F}[s] \cdot \mathcal{F}[r]$. The Fourier transform $\mathcal{F}$ is invertible under certain conditions. We
denote the inverse of \fft by $\mathcal{F}^{-1}$. Finally, we denote the Short-Time Fourier transform (STFT) by
$\mathcal{F}_T$ and its inverse $\mathcal{F}_T^{-1}$. See~\cite{oppenheim2013} for a detailed presentation of the
Fourier transform.

As with all occlusion-based methods, all \freqrex mutations are OOD and are also likely to exhibit at least some degree
of spectral leakage. \freqrex requires that its output have a minimal confidence level. Moreover, we introduce two
stability conditions, one on \emph{causal responsibility} and another on sufficient subset discovery---chain length---in
the paper, to reduce the risk that the output is largely stochastic.
 
\subsection{Actual causality}

We assume that the world is described in terms of variables and their values. Some variables may have a causal influence
on others. This influence is modeled by a set of \emph{structural equations}. It is conceptually useful to split the
variables into two sets: the \emph{exogenous} variables $\U$, whose values are determined by external factors, and the
\emph{endogenous} variables $\V$, whose values are ultimately determined by the exogenous variables. In the causal model
that represents the audio classification, the exogenous variables are those that determine the frequencies present in
the signal. The structural equations $\cF$ describe how these values are determined. A \emph{causal model}, $M$, is
described by its variables, their domains of values, and structural equations. A \emph{context}, $\vec{u}$, is a setting
for the exogenous variables $\U$, which then determines the values of all other variables. In our case, it would
ultimately determine the outcome of classifying the signal with some deep learning model $\mathcal{N}$.

We call a pair $(M,\vec{u})$ consisting of a causal model $M$ and a context $\vec{u}$, a \emph{(causal) setting}. In
order to discover which parts of the signal caused a particular classification $f(s) = o$, we need to make changes
to those elements, removing (or changing) them to assess their effect on the outcome. These changes are called
\emph{interventions}. An intervention is defined as setting the value of some variable $X$ to $x$, and amounts to
replacing the equation for $X$ in $\cF$ by $X = x$.

A Boolean formula $\varphi$ over the set of variables of $M$ is true or false in a causal setting. We write $(M,\vec{u})
\sat \varphi$ if $\varphi$ is true in the setting $(M,\vec{u})$. In this paper, we use the obvious formula to model the
outcome of a classification task as $O \in \{0, 1\}$ where $0$ indicates ``different class'' and $1$ indicates ``same
class''. When all variables are binary, as in this paper, the model is a \emph{binary causal model}.

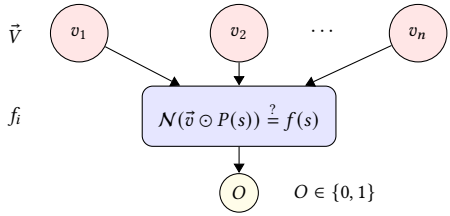
\begin{figure}[t]
    \centering
    \begin{tikzpicture}[outer sep=auto, scale=0.85, transform shape]
        \node (V) at (-2, 0) {$\vec{V}$};
        \node (v1) at (-1, 0) [draw, circle, minimum size=0.9cm,fill=red!10] {$v_1$};
        \node (v2) at (1.5, 0) [draw, circle, minimum size=0.9cm,fill=red!10] {$v_2$};
        \node (dots) at (2.8, 0) {$\mathbf{\cdots}$};
        \node (vn) at (4.3, 0) [draw, circle, minimum size=0.9cm,fill=red!10] {$v_n$};

        \node (F) at (-2, -1.3) {$f_i$};
        \node (f) at (1.5, -1.3)
            [draw, rectangle, rounded corners, fill=blue!10, inner sep=0.25cm]
            {$\mathcal{N}(\vec{v} \odot P(s)) \? f(s)$};

        \draw [-Triangle] (v1) -- (f);
        \draw [-Triangle] (v2) -- (f);
        \draw [-Triangle] (vn) -- (f);

        \node (o) at (1.5, -2.5)
            [draw, circle, minimum size=0.5cm, fill=yellow!10] {$O$};
        \node (oin) at (3, -2.5) {$O \in \{0,1\}$};

        \draw [-Triangle] (f) -- (o);
    \end{tikzpicture}
    \caption{Graphical representation of an audio depth-$2$ causal model with $n$ frequencies as input, and
    classification as outcome. The internal node $f_i$ introduces the network $f$ as a black-box.}
    \label{fig:MXx}
\end{figure}

Given an audio classifier (e.g., a neural network) $f$ and an input signal $s$, we define a binary causal
model $M_{f,s}$ as follows. The set $\V = \vec{V} \cup \{f_i\} \cup \{O\}$ of endogenous variables
consists of a set $\vec{V}$ corresponding to the set of frequencies \fft$(s)$ of $s$, the internal node $f_i$
which incorporates the model $f$ in to the causal model, and the single output variable $O$.

Essentially, $\vec{V}$ is a \emph{mask}, indicating which frequencies of $s$ are present and which are altered, and the
output variable $O$ indicates whether the classification of a partially filtered signal stays the same as of the
original signal. To keep in line with the formal definitions of structural causal models, the values of $\vec{V}$ are
determined by the set of exogenous variables $\U$.

\Cref{fig:MXx} depicts the general structure of the causal model, $\mathcal{M}_{f,s}$, used by~\freqrex, depicting all
endogenous variables. We omit the exogenous variables. Each variable $v \in \vec{V}$ maps directly to one of the
frequencies in the Fourier transform of $s$. The causal model has depth $2$. In what follows, we omit the subscript
${f,s}$ from the causal model notation if it is clear from the context.

This construction assumes causal independence between the variables in $\vec{V}$. Audio classification models perform
classifications over data, not over the original source of the signal in the real world. Naturally, the audio data
encodes correlations, which stem from potential causal connections between the variables in the real world. Disrupting
correlations, however, is precisely how tools such as \freqrex extract knowledge about the neural net. Indeed, obscuring
or altering frequencies is conceptually no different from applying a filter to the signal.
See~\cite{kelly2026iguessthatsblues} for a fuller justification.

Given a classifier model $f$, and input signal $s$, let $\vec{u}_1$ be the context that assigns $1$ to all
variables in $\vec{V}$, and let $\vec{u}_0$ be the context that assigns $0$ to all these variables, where $1$ means the
feature is present with its original value, and $0$ the feature has been removed (set to $0$) or replaced with some
other value. In other words, $\vec{u}_1$ matches the original input exactly, and $\vec{u}_0$ matches the ``empty'' input.
The following definition is for a \emph{minimal sufficient explanation} for the classification.

\dfn[Sufficient Explanation]\label{defn:simple-exp}
A subset $\vec{V}_{exp}$ of $\vec{V}$ is a \emph{sufficient explanation} of the outcome $O = 1$ 
if the following conditions hold:
\begin{description}
\item[{\rm EX1.}] $(M,\vec{u}_0) \models [\vec{V}_{exp} = 1](O=1)$,
where $\vec{V}_{exp} = 1$ stands for assigning $1$ to all variables in 
$\vec{V}_{exp}$.
\item[{\rm EX2.}] $\vec{V}_{exp}$ is minimal; there is no strict subset $\vec{V}_{exp}'$ of $\vec{V}_{exp}$ that satisfies EX1,
  where $\vec{v}'$ is the restriction of $\vec{v}$ to the variables in $\vec{V}_{exp}'$.
\end{description}
\edfn

As there is a one-to-one correspondence between the variables in $\vec{V}$ and the frequencies in $x$, 
$\vec{V}_{exp}$ is one-to-one mapped to a subset of frequencies that are sufficient for the outcome.
In other words, an explanation is a minimal subset of frequencies of a given signal $s$ that is \emph{sufficient} 
for $O=1$. Unfortunately, the precise computation of explanations is intractable~\cite{chockler2024causal}, thus all
explanations are approximate in their minimality.

\subsection{Sufficiency and completeness in audio}
We now present the relevant domain-specific definitions introduced by~\cite{kelly2026guess}. These are the definitions
that we will use the remainder of the paper. 

Given a classifier model $f$, a signal $s$ and its Fourier transform $\mathcal{F}(s)$, a sufficient
subset is: 

\begin{definition}[Sufficient Subset]
    A sufficient subset is a subset $S \subseteq \mathcal{F}(s)$ such that $f(\mathcal{F}^{-1}(S)) =
    f(s)$ and that $S$ is minimal, \ie there is no strict subset $S' \subset S$ such that
    $f(\mathcal{F}^{-1}(S')) = f(s)$.
\end{definition}
\noindent
Here $f(s)$ refers to the top-$1$---the single most probable---class prediction, irrespective of its exact score, softmax or logit. 

When a sufficient subset $S$ is translated into the time domain, it is a \emph{sufficient signal}. 
There is absolutely no requirement that a sufficient signal be unique: indeed it is more likely that a signal
has multiple sufficiencies, as also found in images~\cite{CKK25}.

A \emph{complete subset} is a non-reducible set of frequencies such that they are, by themselves, both \emph{sufficient
and necessary} for the original classification: by themselves they have the required class, and filtering them from $s$
changes the output such that $M \models O=0$. This subset as a \emph{complete signal} when in the time domain. If we
also require some minimal confidence on the explanation, this is a $\delta$-complete explanation, where the explanation
has \emph{at least} $\delta \cdot \sigma(f(s))$ of the original model confidence, $\sigma(f(s))$.

\begin{definition}[$\delta$-Complete Subset]
    A $\delta$-complete subset is a subset $N \subseteq \mathcal{F}(x)$ such that $f(\mathcal{F}^{-1}(N)) =
    f(x)$ and $f(\mathcal{F}^{-1}(\mathcal{F}(x) \setminus N)) \neq f(x)$ and 
    $ \sigma(f(\mathcal{F}^{-1}(N))) \ge  \delta \cdot \sigma(f(x))$
    and that $N$ is minimal, \ie there is no strict subset of $N$ which satisfies this requirement.
\end{definition}

A $\delta$-complete signal is a subset of the original signal which captures its main characteristics as determined by
the model. As we do not further explore the notion of $\delta$-completeness in the paper, we will refer simply to
complete signals unless it is ambiguous.

We require one more concept for the present paper, which is the ``inverse classification''. Given a signal $s$ and its
complete signal $c$, one can sometimes query $f(s - c)$, \ie the ``left over'' frequencies after the complete
signal has been filtered from the original signal. This is not always defined, as the complete signal may in fact be the
entirety of $s$.

\section{Causal Transferability} 
Now that we have the required background, we introduce and formalize our investigation
into the degree to which sufficient and complete signals transfer their classification across models. Essentially, the
question is: \emph{if $s_c$ is a sufficient signal for class $c$ on model $f$, is it also sufficient for $c$ on models
$f_1,f_2, \ldots, f_n$?} To the best of our knowledge, this is the first investigation into the problem in any domain
(see~\Cref{sec:relwork}). 

We now present the definition of transferability for audio signals. Let $s$ be a signal and $f$ be a classifier model
with $n$ classes such that $f(s) = c$. Let $\mathcal{M} = \{ \hat{f}_0, \ldots, \hat{f}_{|\mathcal{M}|-1} \}$ be a set of
models, $\hat{f}$, defined over the same set of classes as $f$. Let $\delta(f(s))$ be a probability distribution over the
classes of $f(s)$. Finally, let $\epsilon$ be a real-valued number such that $0.0 < \epsilon \le 1.0$. We assume
$\log$ to be base $2$.

\begin{definition}[Transferability]
A signal $s$ with class $c$ is \emph{transferable} from $f$ iff for all $\hat{f} \in \mathcal{M}$,
\[ \hat{f}(s) = c \land H(\delta(\hat{f}(s))) \le \epsilon \cdot \log(n),\] 
where $H(\delta(\hat{f}(s)))$ is the Shannon entropy of the output distribution, 
and $\log{n}$ is the maximum entropy for the output of $\hat{f}$. 
\end{definition}
\noindent 

The first part of the definition is clear: a signal is transferable across models if it maintains the same class across
all models. This definition, however, neglects to take into account model confidence. Unfortunately, a simple comparison
of model confidence scores is not usually possible, as different architectures can have remarkably different patterns to
how they distribute confidence. Transformers, for example, typically report much higher confidence than
CNNs~\cite{kelly2025big}. A signal $s$ with high confidence on model $f$ may transfer to $\hat{f}$ but with a score
near the level of noise on that model, \ie the distribution $\delta\hat{f}(s)$ is close to
being flat, indicating almost total uncertainty.

We capture this uncertainty by calculating the maximum entropy over the number of classes of $f$. The maximum entropy for a
categorical distribution, as we have here, is given by $\log(n)$ where $n$ is the number of classes of $f$. From this,
we simply require that $\delta\hat{f}(s)$ be `reasonably' far from flatness, where `reasonably' is controlled by
$\epsilon$. If we set $\epsilon$ to $1$ then we do not care about the probabilities, and only the class will matter.
Setting $\epsilon$ to $0.0$ would prevent all transferability, so we exclude that possibility from the definition.

Given this definition, combining class and confidence, there are two ways in which a signal may be only partially
transferable. Firstly, it may not have the same class on some, or all, of the other models, and secondly, the confidence
of some of the models may be so low as to be too close to noise.

\begin{definition}[Partial Transferability]
  A signal $s$ is partially transferable from model $f$ with goodness ($\alpha, \beta$) relative to $\mathcal{M}$ 
  for $0.0 < \alpha,\beta < 1.0$ if
\[ \alpha = \frac{1}{|\mathcal{M}|} | \{\ \hat{f} | \hat{f}(s) = c \land \hat{f} \in \mathcal{M}\} | \mbox{ and} \] 
\[ \beta =  \frac{1}{|\mathcal{M}|} | \{ \hat{f} | \delta(\hat{f}(s)) \cdot \log(n) \land \hat{f} \in \mathcal{M} \} |.\]
\end{definition}

\section{Transferability Results}

We examined $13$ different models and $4$ data sets. In total, we considered three different classification tasks: i)
emotion recognition from spoken audio; ii) genre recognition from musical audio; and iii) deepfake classification from
spoken audio. As far as possible, we used a variety of different architectures over each dataset to ensure a degree of
model diversity. We used $\delta = 0.5$ when calculating all signal subsets, meaning that both sufficient and complete
signals needed to have at least $50\%$ of the (original) model confidence on the full input signal.

The voice emotion data is taken from the spoken part of the \ravdess~\cite{ravdess} dataset. \ravdess consists of 24
professional actors, equally split between male and female, all speaking with a neutral North American accent. The
different speech emotions are calm, happy, sad, angry, fearful, surprise, and disgust (see~\Cref{fig:disgust}). The
dataset for the music genre classification is \gtzan~\cite{GTZAN}, which is the most popular dataset for audio
classification. This dataset consists of $1000$ audio files in \emph{wav} format. Each file is $30$ seconds long and is
categorized into one of $10$ difference classes: blues, disco, metal, reggae, rock, classical, jazz, hiphop, country or
pop. This dataset has limitations~\cite{sturm2013gtzan}, but we are not concerned directly with the dataset quality, as
we are not training or evaluating models.

For the audio spoof data, we used samples from two popular datasets:
ASVspoof2019~\cite{asvspoof} and ``In The Wild'' (ITW)~\cite{muller2022does}.
As we investigate the spoof datasets in greater depth, we defer those results until~\Cref{sec:spoof}.

The datasets all contain audio recorded at $16$,$000$ samples per second, in mono. The experiments were run on NVIDIA
A100 GPUs, with Ubuntu LTS 22.04. All models are publicly available via
HuggingFace~\cite{wolf-etal-2020-transformers}\footnote{\url{https://huggingface.co/}}. We used $5$ different models on
the \gtzan datasets, $5$ on \ravdess and $3$ on the spoof data. All models are fine-tuned versions of large baseline
audio models (henceforth referred to as backbones). Full details are available in the supplementary material.
We also leave discussion of $\beta$-transferability to the supplementary material, as this varies in line with the degree of entropy permissiveness ($\epsilon$). 

\subsection{Voice Emotion}

\Cref{tab:suff_emotion} shows the degree of $\alpha$-transferability of sufficient subsets and complete
subsets across $5$ different models on the \ravdess voice emotion dataset. Of all the models, \firdhokk most exhibits the
features of a ``flat-earther''. Its sufficiencies are generally not accepted by the other models, and---with the notable
exception of \hondage---it does not readily accept the signals of others. Of particular interest is that, contrary to
all other models, the transferability from \hondage to \firdhokk actually \emph{decreases} when using complete signals.

\begin{table}[t]
    \centering
    \resizebox{\linewidth}{!}{
    \begin{tabular}{l|r|r||r|r|r|r|r|r|r|}
        \multicolumn{3}{c}{} & 
        \multicolumn{2}{||c}{Wav2Vec2} & 
        \multicolumn{1}{|c}{Hubert} & 
        \multicolumn{1}{|c}{DistHubt.} & 
        \multicolumn{1}{|c}{Whisp.} &
        \multicolumn{1}{|c}{} \\ 
        \midrule
        \multicolumn{1}{c}{Model} & 
        \multicolumn{1}{|c}{Size (M)} & 
        \multicolumn{1}{|c}{Acc.} & 
        \multicolumn{1}{||c}{\wiam} &
        \multicolumn{1}{|c}{\hondage} &
        \multicolumn{1}{|c}{\hubert} &
        \multicolumn{1}{|c}{\pollner} &
        \multicolumn{1}{|c}{\firdhokk} &
        \multicolumn{1}{|c}{Avg.} \\
        \midrule
        \wiam     & $86.6$ & $0.89$  & -- & 0.123 & 0.278 & 0.289 & 0.212 & 0.23 \\
        \hondage  & $23.4$ & $0.87$&  0.051 & -- & 0.077 & 0.101 & 0.789 & 0.25 \\
        \hubert& $8.3$  & $0.91$  & 0.17 & 0.073 & -- & 0.184 & 0.189 & 0.15 \\
        \pollner & $23.4$ & $0.9$ & 0.175 & 0.123 & 0.192 & -- & 0.159 & 0.16 \\
        \firdhokk  & $23.4$ & $0.82$ & 0.288 & 0.016 & 0.062 & 0.077 & --&  \textbf{0.11}\\
        \midrule\midrule
        \wiam      & $86.6$ & $0.89$ & -- & 0.133 & 0.383 & 0.337 & 0.324 & 0.29 \\ 
        \hondage   & $23.4$ & $0.87$ & 0.142 & -- & 0.14 & 0.383 & 0.757&  0.36\\
        \hubert    & $8.3$  & $0.91$ & 0.316 & 0.08 & -- & 0.427 & 0.274 & 0.27 \\
        \pollner   & $23.4$ & $0.9$  & 0.276 & 0.115 & 0.377 & -- & 0.214 & 0.25 \\
        \firdhokk  & $23.4$ & $0.82$ & 0.242 & 0.016 & 0.218 & 0.22 & -- & \textbf{0.17}\\
        \bottomrule 
    \end{tabular}
    }
    \caption{$\alpha$-transferability of voice emotion sufficient (top) and complete (bottom) signals across models. Size is in
    millions of parameters. We have also indicated which type of backbone of each model.}
    \label{tab:suff_emotion}
\end{table}

There are small, positive, differences between the transferability rates of sufficient and complete signals on this
dataset. In general, it appears that complete signals are more transferable. To check this, we performed a paired T-Test
on the sufficient and complete results: on these results, however, we were unable to reject the null hypothesis (for $p
= 0.05$) that sufficient and complete transferability are drawn from distributions with different means.

\subsection{Music Genre}

\begin{table}[t]
    \centering
    \resizebox{\linewidth}{!}{
    \begin{tabular}{l|r|r||r|r|r|r|r|r|}
        \multicolumn{3}{c}{} & 
        \multicolumn{3}{||c}{DistHubt} & 
        \multicolumn{1}{|c}{Whisp.} & 
        \multicolumn{1}{|c|}{Trans.} &
        \multicolumn{1}{c}{}\\
        \midrule
        \multicolumn{1}{c}{Model} & 
        \multicolumn{1}{|c}{Size (M)} & 
        \multicolumn{1}{|c}{Acc.} & 
        \multicolumn{1}{||c}{\sanchit} &
        \multicolumn{1}{|c}{\atuaans} &
        \multicolumn{1}{|c}{\pedro} &
        \multicolumn{1}{|c}{\whisper} &
        \multicolumn{1}{|c}{\AST} &
        \multicolumn{1}{c}{Avg.}\\
        \midrule
        \sanchit & $23.7$ & $0.82$  & -- & 0.171 & 0.408 & 0.248 & 0.24 & 0.27 \\
        \atuaans & $94.6$  & $0.86$ & 0.327 & -- & 0.238 & 0.248 & 0.265 & 0.27 \\
        \pedro &$23.5$ &  $0.89$      & 0.47 & 0.218 & -- & 0.253 & 0.19  & 0.28 \\
        \whisper  & $8.31$ & $0.91$ & 0.315 & 0.211 & 0.351 & -- & 0.305 & 0.29 \\
        \AST & $86.6$ & $0.89$      & 0.146 & 0.21 & 0.174 & 0.261 & -- & 0.2 \\
        \midrule\midrule
        \sanchit  & $23.7$& $0.82$ & -- & 0.277 & 0.514 & 0.282 & 0.434 & 0.38 \\
        \atuaans & $94.6$& $0.86$  &  0.384 & -- & 0.449 & 0.295 & 0.339 &  0.37\\
        \pedro  & $23.5$& $0.89$   & 0.573 & 0.338 & -- & 0.297 & 0.4 & 0.4\\
        \whisper & $8.31$& $0.91$  &  0.479 & 0.409 & 0.478 & -- & 0.603 & 0.49\\
        \AST & $86.6$ & $0.89$    & 0.491 & 0.451 & 0.419 & 0.451 & -- & 0.45\\
        \bottomrule 
    \end{tabular}
    }
    \caption{$\alpha$-transferability of music genre sufficient (top) and complete (bottom) signals across models; ``Size'' is in
    millions of parameters. In general, complete signals are more easily transferable across models.}%
    \label{tab:suff_genre}
\end{table}

\Cref{tab:suff_genre} shows the degree of $\alpha$-transferability of genre classification subset signals across the
tested models. In general, \pedro's sufficient signals are the most transferable, with \AST being the least
transferable. ~\Cref{tab:suff_genre} also shows that complete signals are again more transferable than mere
sufficiencies (cf~\Cref{tab:suff_emotion}). In general though, the mean stability of music genre transferability is
greater that for voice emotion. While mean performance is roughly equivalent,~\Cref{tab:suff_genre} reveals large
individual differences between the models.


\section{Spoof Audio}\label{sec:spoof}

\Cref{tab:suff_spoof} shows the results for transferability on spoof audio data from our two datasets, \asvspoof and
\ritw. The transferability rate is much higher on this audio task than for either music or voice emotion recognition.
\melody stands out on \asvspoof by having a much lower rate of transferability, characteristic of `flat-earther' models.
Interestingly, this is no longer the case for the complete signals. This likely indicates a degree of overfitting on the
training data. Indeed, one can see clearly the effects of overfitting by comparing the transferability for \asvspoof (on
which the models were trained) with transferability on ITW.

\begin{table}[t]
    \centering
    \resizebox{\linewidth}{!}{
    \begin{tabular}{c|l|r|r||r|r|r|r|}
        \multicolumn{4}{c}{} & 
        \multicolumn{3}{||c}{Wav2Vec2} &
        \multicolumn{1}{c}{}\\
        \midrule
        \multicolumn{1}{c}{Dataset} & 
        \multicolumn{1}{c}{Model} & 
        \multicolumn{1}{|c}{Size (M)} & 
        \multicolumn{1}{|c}{Acc.} & 
        \multicolumn{1}{||c}{\gust} &
        \multicolumn{1}{|c}{\mo} &
        \multicolumn{1}{|c}{\melody} &
        \multicolumn{1}{|c}{Avg.} \\
        \midrule
                   & \gust & $300$ & $0.93$ & -- & 0.859 & 0.782 & 0.83 \\
                   & \mo & $94.6$ & $0.99$ & 0.192 & -- & 0.882 & 0.54\\
        ASV  & \melody & $94.6$ & $0.99$ & 0.354 & 0.254 & -- & \textbf{0.3}\\
        \cmidrule{2-8}
        Spoof2019 & \gust & $300$ & $0.93$ & -- & 0.764 & 0.664 & 0.71 \\
                  & \mo & $94.6$ & $0.99$ & 0.856 & -- & 0.766 & 0.81 \\
                  & \melody & $94.6$ & $0.99$ & 0.797 & 0.72 & -- & 0.76\\
        \midrule\midrule
                  & \gust & $300$ & $0.93$ & -- & 0.44 & 0.271 & \textbf{0.36}\\
                  & \mo & $94.6$ & $0.99$ & 0.567 & -- & 0.539 & 0.55 \\
        In  & \melody & $94.6$ & $0.99$ & 0.668 & 0.421 & -- & 0.54 \\
            \cmidrule{2-8}
        The Wild & \gust & $300$ & $0.93$ & -- & 0.411 & 0.246 & \textbf{0.33}\\
                & \mo & $94.6$ & $0.99$ & 0.679 & -- & 0.521 & 0.6\\
                & \melody & $94.6$ & $0.99$ & 0.71 & 0.517 & -- & 0.61\\
        \midrule
    \end{tabular}
  }
    \caption{$\alpha$-transferability of spoof sufficient (top) and complete (bottom) signals across models on ASVspoof2019 and
      ITW. ``Size'' is in millions of parameters. Highlighted are average transferability rates which are unusual.}
    \label{tab:suff_spoof}
\end{table}

\subsection{Spoof Global Composition}\label{subsec:spoof_comp}

\cite{kelly2026guess} introduce a simple procedure where they compose together as many sufficient signals, $s_s$, for a
model $f$ with the same classification $c$ as possible, such that $\kappa = \{ s_s | f(s_s) = c\}$ is the largest set
that satisfies $f(\frac{1}{|\kappa|} \sum_{k \in \kappa} \kappa) = c$. In that paper, the aim was to show that the
models were not relying on meaningful musical cues, as the resultant compositions---as many as $100$ different
sufficient signals superimposed---were still classified as the original label but were audibly far removed from
``blues''. In this paper, we suggest it as a reasonable proxy for a global sufficient ``signature''. This does not imply
that it is a unique global signature, of course: it is perfectly reasonable that a classification have more
than one independent reason to occur.

\begin{table}[t]
    \begin{tabular}{c||r|r||r|r}
      \multicolumn{1}{c}{} & \multicolumn{2}{||c||}{ASVspoof} & \multicolumn{2}{c}{ITW} \\
      \toprule
      \multicolumn{1}{c}{} & \multicolumn{2}{||c||}{Class} & \multicolumn{2}{||c}{Class} \\
      \midrule\midrule
      Model & real & fake & real & fake \\
      \midrule
      \gust   & 0.996 & 0.47 & 1.0 & 0.64 \\ 
      \mo & 0.6 & 0.97 & 0.57 & 0.95 \\
      \melody & 0.47 & 0.99 & 0.52 & 1.0 \\
      \bottomrule
    \end{tabular}
    \caption{Degree of compositionality for real and fake classifications across 3 models and ASVspoof2019 (left) and ITW
    (right).}%
    \label{tab:spoof_comp}
\end{table}

\Cref{tab:spoof_comp} shows the degree of compositionality of real and fake audio for $3$ different models over two
different datasets, ASVspoof2019 and ITW. All three models were fine-tuned on \asvspoof, so the ITW data is potentially
out of distribution (OOD). The compositional behavior of \gust is different from the other two models, indicating that
it might be a ``flat-earther'' model. In particular, \gust{'s} compositionality relationship between real and fake is
inverted compared to the other models, with real being much easier to compose.

\begin{figure*}[t]
    \centering
    \includegraphics[scale=0.65]{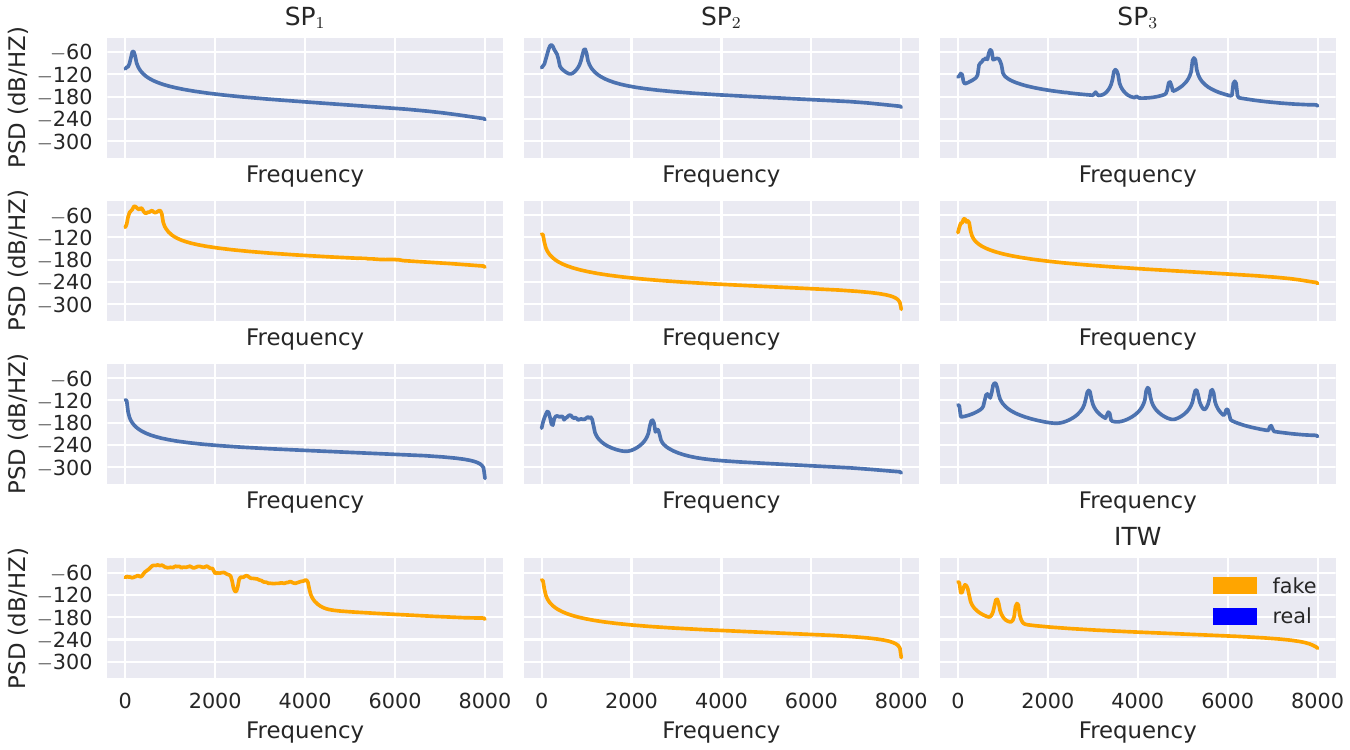}
    \caption{Power spectral density of \textcolor{blue}{real} and \textcolor{orange}{fake} data on \asvspoof (top $2$ rows) and
    ITW (bottom two rows). The different between \gust and the other models in clear.}%
    \label{fig:real_fake}
\end{figure*}

\Cref{fig:real_fake} shows the difference in the power spectral densities of composite signals between real and fake
across $3$ different models on ASVspoof2019 (top two rows) and ITW (bottom two rows). It reveals an interesting
divergence between the models. \gust requires very fewer frequencies in the real category, while fake exhibits more noise.
This pattern is reversed however for the other two models, where real shows considerably more peaks than fake, which is
very smooth. Regardless, the pattern is still clear: for all models, there is a clear difference between the global
signatures of real and fake classifications, though the direction of the pattern is not consistent.

\subsection{Spoof Entropy}\label{subsec:entropy}

\begin{figure*}
  \centering 
  \includegraphics[scale=0.70]{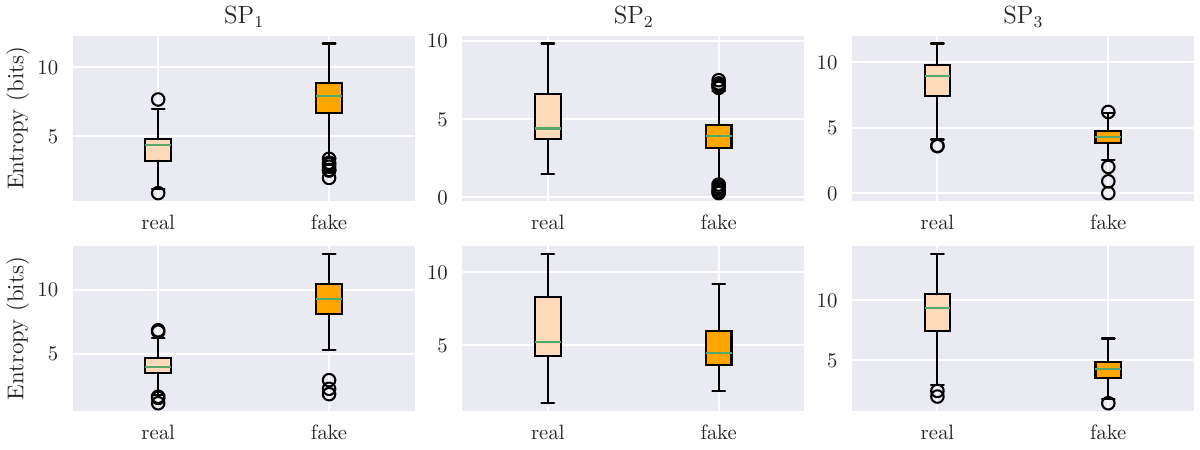}
  \caption{Spectral entropy across real and fake sufficiencies on \asvspoof (top) and ITW (bottom) and $3$ models.}
  \label{fig:spoof_entropy}
\end{figure*}

\Cref{fig:spoof_entropy} shows that, across models and datasets, there is always a significant difference in spectral
entropy across the `real' and `fake' classes. Unfortunately, however, it is not unidirectional
(cf~\Cref{fig:real_fake}). \gust is an outlier among the $3$ models as audio classified as `real' has lower spectral
entropy than `fake'. This is the opposite pattern to the other $2$ models. We performed a Mann Whitney U Test with the
null hypothesis that there was no difference in entropy between real and fake: this was strongly rejected by the
statistical test.

\subsection{Transferability across labels}

We performed an additional experiment to see if we could transfer the global signature for ``fake'' onto data classified
as real. For this we took the sufficient signal from the ``fake'' composite, as detailed in~\Cref{subsec:spoof_comp},
and applied it to real data. We made no other alterations to the signal. On \asvspoof, it was possible to change real to
fake on \melody $53\%$ of the time, on \mo $50\%$ and $0\%$ on \gust. It was not possible to convert real to fake using
this procedure. The pattern is similar for the ITW dataset: \melody $53\%$, \mo $52\%$ and just $3\%$ on \gust.

This result further strengthens the argument that \gust is an outlier model---a flat-earther---when compared to the
other $2$, and this despite having the same backbone: a Wav2Vec2 model. Even though it boasts a similar accuracy to the
other models, $93\%$ vs $99\%$, analysis of its sufficient and complete signals demonstrate that it is using different,
though still valid, frequencies from the other $2$ models.

\subsection{Automatic Speech Recognition}\label{subsec:asr}

\begin{table}[t]
    \centering
    \begin{tabular}{c|r|r|r|r|r|r}
        \multicolumn{1}{c}{Model} &
        \multicolumn{2}{|c}{Sufficient} &
        \multicolumn{2}{|c}{Complete} &
        \multicolumn{2}{|c}{Inverse STOI} \\
        \midrule
         & real & fake & real & fake & real & fake \\
        \midrule
        \gust  & $0.99$ & $0.99$ & $0.99$ & $0.98$ & $0.14$ & $0.15$\\
        \mo & $0.99$ & $0.98$ & $0.99$ & $0.98$ & $0.05$ & $0.06$\\
        \melody & $0.99$ & $0.98$ & $0.99$ & $0.98$ & $0.02$ & $0.01$\\
        \midrule\midrule
        \gust  & $0.99$ & $0.99$ & $0.99$ & $0.98$ & $0.58$ & $0.63$\\
        \mo & $0.99$ & $0.99$ & $0.99$ & $0.98$ & $0.03$ & $0.07$\\
        \melody & $0.99$ & $0.99$ & $0.99$ & $0.98$ & $0.05$ & $0.02$\\
        \bottomrule
    \end{tabular}
    \caption{Levenshtein ratio of sufficient and complete signals using an ASR model as oracle. While all models show
    virtually identical fidelity on sufficient and complete signals, there is considerable difference on the intelligibility
  of the \emph{inverse} signal: the signal remaining after the complete signal has been removed. \gust is again an
outlier, which a great deal of intelligible signal remaining, unlike the other models.}
    \label{tab:avs_distance}
\end{table}

For the spoof data, we used an automatic speech recognition (ASR) model to produce text from the original audio,
sufficient audio, complete audio and inverse audio. We calculated the Levenshtein
distance~\cite{Levenshtein1965BinaryCC} between the transcript of the original audio, and the transcriptions of the
subset signals. The Levenshtein distance between two strings, $s_1$ and $s_2$, is the minimal edit distance to turn
$s_1$ into $s_2$. From this we calculated the Levenshtein ratio: how much of the string remains the same or is altered,
with $1$ indicating no changes were required to turn $s_1$ into $s_2$ (\ie they are already equal), and $0$ indicating
complete dissimilarity between the two strings.

We also calculated the Short Term Objective Intelligibility measure (STOI)~\cite{stoi}
between the original audio and the various sub-signals derived from it. In particular, we present the results for STOI
on the inverse audio. The inverse audio is the signal derived from the subset of frequencies \emph{not} used in the
complete subset. In general, one would likely expect these frequencies to be outside the normal vocal range, assuming
that the complete signal focused on voice. We present the results in~\Cref{tab:avs_distance}.

In terms of string edit distance, there is no statistically significant difference between sufficient and complete
signals on all models. The Levenshtein ratio is always $\approx 0.99$, indicating a high degree of fidelity between the
original text and the text derived from subset signals. Sufficient and complete subsets are not usually the same size:
this is a good indicator that, although sufficient and complete signals are capturing essentially the same audible
signal, the models are all using other frequencies, possibly non-audible, in their complete signals.

Of more interest is the STOI of the inverse classification. The inverse signal is created by settings all frequencies in
the complete signal to $0$: it is the ``left over'' which the model does not require for its classification. For both
\mo and \melody, there is very little voice-range signal left over once the complete signal has been removed. The
complete signal for both of these models captures nearly all the frequencies associated with speech. This can be seen in
the uniformly low STOI, though there is a difference between `real' and `fake' classifications. Again, `fake' has
slightly more spectral information in the speech range than `real', and this has a tendency to be leaked by the model.
\gust, our `flat-earther` shows very different STOI behavior, with a great deal of audible speech remaining in the
inverse signal, especially for `fake'. This again indicates that, despite having the same backbone and training data,
other factors are at play in model performance.

\section{Discussion}

The degree of transferability varies between the spoof task on one hand, and emotion and music genre recognition on the
other. One might suspect that the number of classes would influence the degree of transferability, but this does not
seem to be the case. Voice emotion recognition has the lowest degree of transferability, even though it has more classes
than spoof and fewer than music genre. These models have lower accuracy in general than the models on the others tasks.
Moreover, it does not seem to be the case that signals are more transferable across models sharing the same backbone as
opposed to models with different backbones. While models are seemingly using many of the same features in their decision
processes, the overlap is by no means total. Indeed, if models were accurately using music cues, for example, one would
expect to see much higher rates of transferability, and not the range of $20\%-29\%$ that we see here
(\Cref{tab:suff_genre}). The much higher rates of transferability for spoof data might suggest that the models are using
genuine information, rather than shortcuts, but this is clearly not the case for all the models.

\Cref{fig:real_fake} suggests that there is, on average, a difference between the Fourier Transforms (FFT) of real audio and
fake audio. This is in line with work such as~\cite{schafer2025}, which performed a manual analysis in order to discover
deepfake FFT artifacts. Our results suggest that some models are also likely learning FFT specific signatures of spoofed audio. The differences between \asvspoof and ITW suggest that these models are overly specific to their training data~\cite{Mller2022DoesAD}.

\Cref{tab:avs_distance} raises the question: ``what do we want from our models?''. All the models have similar
performance, but the way they consume information (and OOD signals) is clearly different. We find that all $3$ models
accept near silence (from the human perspective) as `real'. This raises the ontological question of whether silence
should count as real or not. Silence can, of course, be real --coming from either natural source--but it may just be an
array of $0$s.

Finally, we note that there does appear to be a weak correlation between size and accuracy of the model and the degree
to which its sufficient signals are transferable. In particular, \gust and \firdhokk have the highest, and lowest,
average transferability in their tasks, while both having the lowest accuracy. Clearly, the relationship is not simple
correlation and points at some other cause for this behavior. Further exploration is needed to understand why some
models behave in a superficially similar fashion to others, yet have fundamentally different properties.

\section{Future Work} We note that transferability rates are broadly in line with black-box adversarial attack
transferability~\cite{success2021}. This opens us the intriguing possibility that there is a connection between
sufficient transferability and adversarial attacks. This investigation has only looks at single sufficient signals per
input and model. We know from images~\cite{CKK25} that inputs often have multiple, independence sufficiencies. It is
likely the case that the transferability rates reported here would be different if we had multiple different signals to
test.

There is increasing evidence to suggest that generative models fall back on a few prototypes~\cite{hintze2026}. If this
is also true of generative AI in audio, then we would expect to see less information in fake signals and this is indeed
what we see (\Cref{fig:spoof_entropy}) for $2$ of the models. This observation warrants further work.

There are models specifically built, ground up, for deepfake detection, rather than being fine-tuned versions of generalist
models~\cite{wang2025audiodeepfakeverification}. For example, \cite{warren2025} develop a prosodic model for deepfake
detection. It would be very interesting to see whether sufficient signals for other deepfake detection models, such as
those used here, would be transferable to a model which is build from an entirely different starting point.

\section{Related Work}\label{sec:relwork}

Transferability has been widely studied from the point of view of adversarial
attacks~\cite{demontis2019,zhao2021success,wang2023beyond}. The aim is to build an attack using a surrogate model
assumed to accurately reflect some aspects of the target model, before moving the attack to the target model. To the
best of our knowledge, nobody has looked at whether sufficient and complete signals are transferable across models and
architectures. There are other approaches to adversarial attacks in
audio~\cite{carlini2018,du2020sirenattack}.~\cite{taori2019targeted} use a black-box method to achieve an attack success rate of $\approx 35\%$, similar to our transferability rates.~\cite{farooq} propose a GAN‑based framework generating attacks that transfer across deepfake detectors; their results show that SOTA detectors lose substantial accuracy under black‑box attacks.

\cite{lopez2024} investigate the transferability of explanations between training and test data, when there are minor changes in distribution. This is a related, but different question from ours, where we ask whether sufficient signals are transferable across models. They do not consider either sufficiency or necessity in their work.

We use \freqrex to build our sufficient and complete signals. Sufficiency in particular is well studied in tabular data,
computer vision~\cite{INMS19,chockler2024causal,darwiche2020} and even robotics~\cite{yaacov2026icra}. We are not aware
of another tool, however, that allows us to do this for audio. There are a number of \xai tools intended to explain
model decisions specifically for audio~\cite{theissler,siglime,mishra2017local,akman2024audio,melchiorre2021lemons}. None of them directly produce output suitable for this investigation.
AudioLime~\cite{haunschmid2020audiolime} and MusicLime~\cite{sotirou2025} produce listenable output but rely on source
separation, rather that FFT-based analysis. The output has no guarantee of minimality. Moreover, neither of these tools
is capable of producing (minimal) sufficient or complete signals. Moreover, these methods will not work on \ravdess, as
there is only one source in the signal. Ultimately, these tools are aimed at human interpretability, which is not what
we require in our analysis.

\cite{sturm2013classification} argues that the classification task is not valid when it comes to making a conclusion
about whether a system is recognizing genre, emotion, or some other semantic category. Our investigation does not seek
to address this question, though it could be extended to address this issue. The problem of what information a model uses
exists in image classification also: classifiers rely on very few pixels (e.g., $2\%$ of the input for
transformers) in order to make their decisions~\cite{kelly2025big}.

\cite{schafer2025} build a dataset of real audio and audio created using SOTA generative methods. Various audio features
are then compared over the real and fake audio which reveals that there are important differences between real and fake
audio. This analysis may serve as a good basis for model training, but does not help to identify which features a model
is actually using. Our results support the same narrative, that there are meaningful differences in the frequencies of
real and fake audio and that classifier models may be detecting them. \cite{Afchar2025AFE} also use an FFT approach to
identify artifacts associated with generative AI. They do not, however, investigate whether such artifacts are
transferable across models.

For images,~\cite{navaratnarajah2025out} show that object detectors are suspectible to small perturbations to
their minimal sufficient pixel sets (MSPSs), but do not explore whether their attacks transfer to
different models. It seems plausible that, if a sufficient signal is transferable, then a successful attack
on that sufficient signal is also likely to transfer.

It is well known that choices outside of architecture and training data can have surprising and profound effects of
model behavior~\cite{dodge2020,sellam2022multiberts}. 
For example,~\cite{picard2021} shows that model accuracy can vary
depending on initial random seed. Our flat-earther models are likely the result of some aspect of training, such as
initial train/test split or random seed.

\section{Conclusion}

We have presented the first investigation into the transferability of subsets of audio signals across different
classifier models. We have provided formal definitions for transferability of audio signals.
We find the degree to which signals are transferable is linked to the classification task. We have
identified that some models have distinct transferability behavior which is not otherwise detected by standard metrics.
We have also performed an investigation into the information theoretic characteristics of deepfake audio, as seen
through the lens of models, and revealed that classifications for real and fake exhibit different characteristics.

\section*{Acknowledgements}
Hana Chockler and David A. Kelly acknowledge support of the UKRI AI program and the Engineering and 
Physical Sciences Research Council 
for CHAI -- Causality in Healthcare AI Hub [grant number EP/Y028856/1].

\bibliographystyle{IEEEtran}
\bibliography{all}

\end{document}